\documentclass{emulateapj}

\usepackage{amssymb}
\usepackage{amsmath}
\usepackage{graphicx}

\newcommand{\f}     {\frac}
\newcommand{\lAD}   {\lambda_{AD}}

\begin{document}

\title{Theoretical expectations for magnetic field measurements in 
       molecular cloud cores}

\author{Fabian Heitsch\altaffilmark{1,2}}
\author{P.S. Li\altaffilmark{3}}
\altaffiltext{1}{Universit\"ats-Sternwarte M\"unchen, Scheinerstr. 1,\\
                 81679 M\"unchen, Germany.
                 email: heitsch@usm.uni-muenchen.de}
\altaffiltext{2}{U Wisconsin-Madison, 475 N Charter St, Madison,
                 WI 53706, U.S.A.}
\altaffiltext{3}{Astronomy Department, University of California at Berkeley,\\
    601 Campbell Hall, Berkeley, CA 94720-3411, USA}

\lefthead{Heitsch \& Li} 
\righthead{Measuring magnetic fields in cores}

\begin{abstract}  
  The currently most viable methods to estimate magnetic field strengths in molecular cloud
  cores are Zeeman measurements and the Chandrasekhar-Fermi (CF) method.  
  The CF-method estimates magnetic field strengths
  from polarimetry and relies on equipartition between the turbulent
  kinetic and turbulent magnetic energy in the observed region. Thus, its application
  to objects not dominated by hydromagnetic turbulence is questionable. We calibrate the CF-method
  and Zeeman measurements against numerical models of self-gravitating molecular cloud cores.
  We find -- in agreement with previous results -- that the CF-method is on the average 
  accurate up to a factor of $2$, while a single estimate can be off by a factor
  of $7$. The CF-method is surprisingly robust even when applied to regions not dominated
  by hydromagnetic turbulence and in the face of small-number statistics.
  Zeeman measurements can systematically overestimate the field strength in high-density regions,
  if the field is predominantly oriented along the line of sight, while they tend to underestimate
  the field in other cases.
\end{abstract}

%
%
\section{Motivation}\label{s:motivation}
Magnetic fields are omnipresent in the interstellar medium (ISM). Their 
importance depends on their strength. The large-scale Galactic magnetic 
field contributes to the scale height of the Galactic disk, and it confines cosmic
rays to the Galaxy, thus influencing the chemistry in molecular clouds.
Molecular cloud cores -- thought to be gravitationally unstable without an additional
support mechanism -- have been supposed to be stabilized by magnetic fields,
leading to the model of magnetically mediated (low mass) star formation 
\citep{SAL1987}. An alternative -- although not necessarily exclusively so 
\citep{LIN2004} -- model pictures magnetic fields as dynamically unimportant,
and star formation being controlled mainly by turbulence on dynamical time 
scales (e.g. Larson 1981; Elmegreen 2000; Hartmann et al. 2001; Mac Low \& Klessen 2004).

Magnetic field strength estimates in molecular cloud cores thus might 
provide a way to distinguish between those two models.  
Such measurements rely mostly on 
the Zeeman effect (e.g. Crutcher 1999, Bourke et al. 2001, Sarma et al. 2002)
resulting in an
estimate of the density-weighted, direction-sensitive integrated magnetic field strength 
along the line of sight, and on the Chandrasekhar-Fermi method \citep{CHF1953}. 
The CF-method relates the line of sight velocity dispersion to the plane of sky polarization 
angle dispersion under the assumption of an isotropically turbulent medium
whose turbulent kinetic and turbulent magnetic energy components are in 
equipartition, yielding an estimate of the mean\footnote{There is a difference
between the mean field with respect to direction and the mean field with respect
to orientation. The former refers to a local field perturbation (which could
be perpendicular to the mean field w.r.t. orientation), while the latter is an
averaged mean field. It is the latter we are referring to.} field
\begin{equation}
  \langle B\rangle^2 = 4\pi\rho\f{\sigma(v)^2}{\sigma(\theta)^2}.
  \label{e:cforig}
\end{equation}
In its original form as given by \citet{CHF1953}, equation~(\ref{e:cforig}) 
assumes small angles around the (required) mean field direction.
Calibrations of the method with the help of numerical simulations of
magnetized turbulent molecular cloud regions show that the CF-method
on average overestimates the ``true'' field strength by a factor
of approximately $2$ 
(Ostriker et al. 2001; Padoan et al. 2001; Heitsch et al. 2001b, in the following
HZMLN). Limited observational
resolution (mimicked by smoothing the simulated polarization maps with a
Gaussian) leads to a further systematic overestimate (HZMLN).
While the method is reasonably robust when selecting for small angles
\citep{OSG2001}, it fails utterly for vanishing mean fields in the plane 
of sky, as expected (HZMLN). 
The latter authors suggested a modification of the
CF-method which estimates not the mean field $\langle B\rangle$, but the rms
field $\langle B^2\rangle^{1/2}$. 

The CF-method complements Zeeman estimates to some extent, since it
infers the mean field strength in the plane of sky. The hope is to 
actually retrieve information about the 3D structure of the magnetic field
\citep{LGC2003}. Many of the more recent
applications of the CF-method refer to the dense phases of the ISM, mostly
to molecular cloud cores or Bok globules 
(e.g. Lai et al. 2001; Henning et al. 2001; Lai et al. 2002; Matthews et al. 2002; 
Lai et al. 2003, Wolf et al. 2003; Crutcher et al. 2004). 
\citet{VGF2003} decided not to use the CF-method because of the uncertainty introduced
by the line-of-sight averaging of the polarization pattern.
Since the method relies on equipartition between
the turbulent kinetic and magnetic energy, its application to other regimes 
than those dominated by hydromagnetic turbulence might raise some questions that we will
discuss in \S\ref{s:cfsystem}.

The goal of this study is twofold: First we aim at establishing 
how reliable the CF-estimates are for regions which are not dominated by 
hydromagnetic turbulence. We calibrate the method
with the help of cores in a high resolution model of self-gravitating
magnetized turbulence \citep{LNM2004}. We find that the CF-method returns
on average reliable estimates (within a factor of $2$) as long as the full 
angle information is used. Single estimates can be off by a factor of $7$. 
The method is surprisingly robust even when it is applied to rotating and
self-gravitating cores. Despite its robustness, the method is subject to various
systematic effects. 

Second, we use the model cores to infer Zeeman measurements of the field strengths
and to compare those to the CF-estimates. 
While Zeeman measurements are expected to underestimate the 
magnetic field in a turbulent region because of cancellation along the line of sight,
we find that they can systematically overestimate the field in high-density regions with ordered
fields oriented parallel to the line of sight.

%
%
\section{The CF-method in cores: systematic effects}\label{s:cfsystem}

The CF-method in its original form rests on four assumptions:
(a)~The field perturbations are caused by hydromagnetic turbulence. 
(b)~The turbulent kinetic and turbulent magnetic energy are in equipartition
(This is guaranteed for Alfv\'{e}nic perturbations, see e.g. \citet{ZWM1995}.). 
(c)~The turbulence is isotropic. (d)~The field is perfectly frozen to the (total)
gas density. 
If one or more of these assumptions are violated, systematic effects will be 
introduced in the CF-estimates (see e.g. Zweibel (1990), 
Myers \& Goodmann (1991), Zweibel (1996) and Crutcher et al. (2004) for
a discussion). In the context of protostellar cores, two effects would be expected
to play a major role: 

(i) The field perturbations are caused by other agents than
turbulence, namely self-gravitation and/or rotation. 
First, we could imagine a core within the picture of magnetically dominated star formation:
The core is permeated by a strong field, rendering
the system subcritical (e.g. \citet{MOS1976}). Turbulence is assumed to be negligible.
The field would be expected to evolve to an hourglass-like structure, which the 
CF-method would in turn interpret as an effect of turbulence, thus underestimating the
field strength.

Second, let us consider a flattened rotating core seen face on such, that the rotation
axis is more or less parallel to the line of sight. The mean magnetic field vector is
oriented parallel to the line of sight (This is the situation envisaged in the concept
of magnetic braking of cores, see e.g. \citet{LUS1955}; \citet{MES1959}). 
The field lines would wind up, i.e. in 
projection they would describe spirals. These would result in large polarization angle
dispersions not necessarily accompanied by a large velocity dispersion along the line
of sight.

(ii) For anisotropic turbulence (see e.g. \citet{CHL2003}), relating the line of sight
velocity dispersion to the polarization angle dispersion can introduce a systematic 
effect in the CF-estimate. However, if the magnetic field is predominantly in the plane
of sky, the CF-method could still yield reasonable results despite severe anisotropy.

 




%
%
\section{Models and Methods}\label{s:modelsnmethods}
The model this study rests on is described in detail
by \citet{LNM2004}. It mimics a fraction of an isothermal
molecular cloud region (periodic boundary conditions), which is
dominated by supersonic turbulence driven on the largest scales
at Mach $10$ such that the driving energy input is kept constant
(for a detailed description see \citet{MAC1999}). Once an equilibrium
state between driving energy input and dissipation has been reached, 
self-gravity is switched on. The shock-generated filaments start
to fragment and form cores, the objects of this study. The initial
ratio of thermal to magnetic pressure is $\beta = 0.9$. For a detailed
discussion of this type of simulation and the choice of parameters 
see e.g. \citet{HMK2001}. 

Cores selected for this study have to fulfill several criteria:
(a) They are gravitationally bound. (b) They are
resolved according to the so-called Jeans-criterion \citep{TKM1997}, given by 
\begin{equation}
  \lambda_J \equiv \sqrt{\f{\pi}{G\rho}}c_s > 4\Delta x,
  \label{e:jeanslength}
\end{equation}
where $\Delta x$ is the grid spacing and $c_s$ is the sound speed.
(c) Regions of surface densities of $10$\% of the peak surface
density or above are used for analysis. Although this (excitation-) threshold 
is chosen somewhat arbitrarily (albeit guided by observational maps),
other choices (within reasonable limits) did not affect the results.
(d) Selected cores consist of more than $1000$ cells in order to
sample their internal structure sufficiently. The sample cores are
listed in Table~\ref{t:cores}. The fourth column in the table denotes
unresolved cores according to the Jeans criterion. These cores are not used for 
the subsequent analysis except to demonstrate resolution effects.
Note that $\rho/\rho_J$ gives a resolution criterion, while $M/M_J$ determines
whether a core is gravitationally unstable. Here, $M_J$ is the thermal Jeans mass.
\begin{deluxetable}{ccccccccc}
  \tablewidth{0pt}
  \tablecaption{Properties of sample cores\label{t:cores}}
  \tablecomments{First column: core label: d{\em i}c{\em j}, where $i$ denotes the time dump
                 number and $j$ specifies the core number at time dump $i$. Second column: maximum
                 over Jeans density, derived from the stability
                 criterion following \citet{TKM1997}. 
                 Unresolved cores
                 are marked with an X in the fourth column.
                 Third column: number of cells within core. 
                 Fifth column: mass contained in core over Jeans mass,
                 assuming a spherical gas distribution.
                 Sixth column:
                 Mach number derived from (density-weighted) line-of-sight 
                 velocity dispersion. 
                 Seventh column: plasma beta $\beta=2\bar{\rho}c_s^2/B^2$.
                 Eighth column: Alfv\'{e}n Mach number ${\cal M}_A = v_{rms}/c_A$
                 where the Alfv\'{e}n speed $c_A=B/\sqrt{4\pi\rho}$.
                 Ninth column: dump time in units of the global free fall time
                 $\tau_{ff}=\sqrt{3\pi/(32 G\rho)}\approx 10^6$yrs.}
  \tablehead{\colhead{core}&\colhead{$\rho/\rho_J$}
             &\colhead{\# cells}&\colhead{}&\colhead{$M/M_J$}&\colhead{$\cal{M}$}
             &\colhead{$\beta$}&\colhead{${\cal M}_A$}&\colhead{$t/\tau_{ff}$}}
  \startdata
  d04c04 & $0.32$ & 5956 &     &$1.4$& $0.2$ &$2.2$ &$0.2$ &$0.50$\\
  d05c02 & $0.33$ & 5283 &     &$1.3$& $0.3$ &$1.9$ &$0.3$ &$0.67$\\
  d06c11 & $0.28$ & 3589 &     &$1.6$& $0.5$ &$3.7$ &$0.7$ &$0.83$\\
  d06c15 & $0.22$ & 2928 &     &$0.8$& $0.5$ &$1.1$ &$0.4$ &$0.83$\\
  d07c04 & $5.68$ & 2167 &  X  &$5.4$& $1.0$ &$2.0$ &$1.0$ &$1.00$\\
  d07c05 & $2.44$ & 1194 &  X  &$2.7$& $0.7$ &$2.3$ &$0.8$ &$1.00$\\
  d07c11 & $0.55$ & 1574 &     &$1.8$& $0.6$ &$0.8$ &$0.4$ &$1.00$\\
  d08c01 & $70.6$ & 2043 &  X  &$23$ & $1.4$ &$0.7$ &$0.8$ &$1.17$\\
  d08c02 & $57.6$ & 2056 &  X  &$21$ & $1.2$ &$1.9$ &$1.2$ &$1.17$\\
  d08c09 & $9.58$ & 1133 &  X  &$6.5$& $1.0$ &$1.6$ &$0.9$ &$1.17$\\
  d08c17 & $0.26$ & 2344 &     &$2.2$& $0.3$ &$6.8$ &$0.6$ &$1.17$\\
  d08c19 & $0.26$ & 1354 &     &$0.5$& $0.3$ &$8.1$ &$0.6$ &$1.17$\\
  d09c14 & $0.91$ & 1440 &     &$1.8$& $0.4$ &$1.5$ &$0.4$ &$1.33$\\
  d09c19 & $0.35$ & 1130 &     &$1.2$& $0.2$ &$3.3$ &$0.3$ &$1.33$
  \enddata
\end{deluxetable}

Summarizing Table~\ref{t:cores}, the model cores are marginally gravitationally bound,
their internal velocities are (in most cases) subsonic (${\cal M} \approx 0.5$), and 
the magnetic pressure is on the order of the thermal pressure (with a substantial 
scatter). 

Once cores have been identified over the run of the simulation, 
we construct surface density maps and
polarization maps following \citet{ZWE1996} via
\begin{equation}
  P = Q + iU = N \int f(y) \frac{(B_x + i B_z)^2}{B_x^2 + B_z^2}\,
      \cos^2\gamma\,dy
  \label{e:stokesparm}
\end{equation}
where $B_x$ and $B_z$ give the field vectors in the plane of sky perpendicular
to the line of sight and are taken directly from the simulations. We integrate
along the line of sight in $y$-direction. 
The function $f(y)$ is a weighting function which accounts for the density,
emissivity, and polarizing properties of the dust grains. As the simplest 
assumption, we take $f(y)$ to be the local gas density normalized by the 
column density $1/N=\int f(y)\,dy$, where we take the opportunity to correct
a typo in HZMLN. 
The factor
\begin{equation}
  \cos^2 \gamma = \frac{B_x^2+B_z^2}{B_x^2+B_y^2+B_z^2}
  \label{e:cosgamma}
\end{equation}
accounts for suppression of polarization by the magnetic field component 
along the line-of-sight (e.g. \citet{FPC2000}, \citet{PGD2001}).
The polarized intensity is $|P| = \sqrt{Q^2 + U^2}$, and the polarization 
angle is
\begin{equation}
  \phi = \frac{1}{2}\arctan{\frac{U}{Q}}.
  \label{e:polarangle}
\end{equation}
Equation~(\ref{e:stokesparm}) is an approximate solution of the full 
radiative transfer equation for the Stokes parameters, 
(Martin 1974; Lee \& Draine 1985),
valid for small polarization and low optical depth. At far infrared
wavelengths, and at typical column densities for molecular clouds, the
medium can safely be assumed to be optically thin \citep{HDD2000}.
We deliberately leave out any statement about the polarization degree as
we are only interested in the polarization angles, but not in the ratio between
polarized and continuum intensity. By this we intend to keep the argument as
simple as possible, not embarking on a discussion on the efficiency of
grain alignment mechanisms (see e.g. \citet{LAZ2003}). 
This simplification certainly limits our study to
the investigation of polarization angles only.
To mimic limited telescope resolution, we smooth the polarization maps
with a Gaussian filter as in HZMLN.

The CF-method is applied to the ``polarization maps'' in the 
extended form given by
\begin{equation}
  \langle B\rangle^2 = 4\pi\rho\f{\sigma(v)^2}{\sigma(\tan\theta)^2},
  \label{e:cfextend}
\end{equation}
where we replace the small-angle approximation by the $\tan$ of the
angle. The density is determined from the column densities of the 
polarization maps, assuming a spherical mass distribution, where
the radius is given by the geometric mean of the long and short
axis of the core \citep{CNW2004}.

To infer a ``Zeeman'' estimate from our cores, we integrate 
\begin{equation}
  \langle B_{ZM}\rangle = N \int f(y) B_y\,dy
  \label{e:zeeman}
\end{equation}
along the line of sight. $N$ and $f(y)$ are the same as in 
equation~(\ref{e:stokesparm}).

%
%
\section{Results}\label{s:results}

\begin{figure}
  \plotone{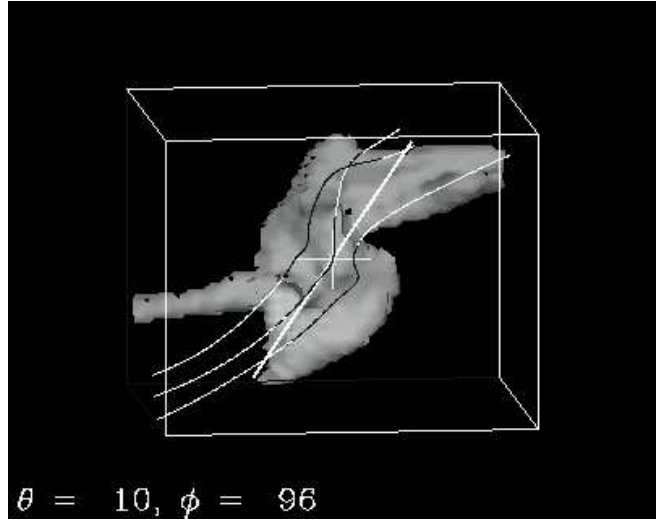}
  \caption{\label{f:core3d}3D core projection. The core outline
           is given by iso-density surfaces. The straight line denotes the
           mean flux direction. Three representative field lines are outlined.}
\end{figure}

\begin{figure*}
  \includegraphics[width=0.32\textwidth]{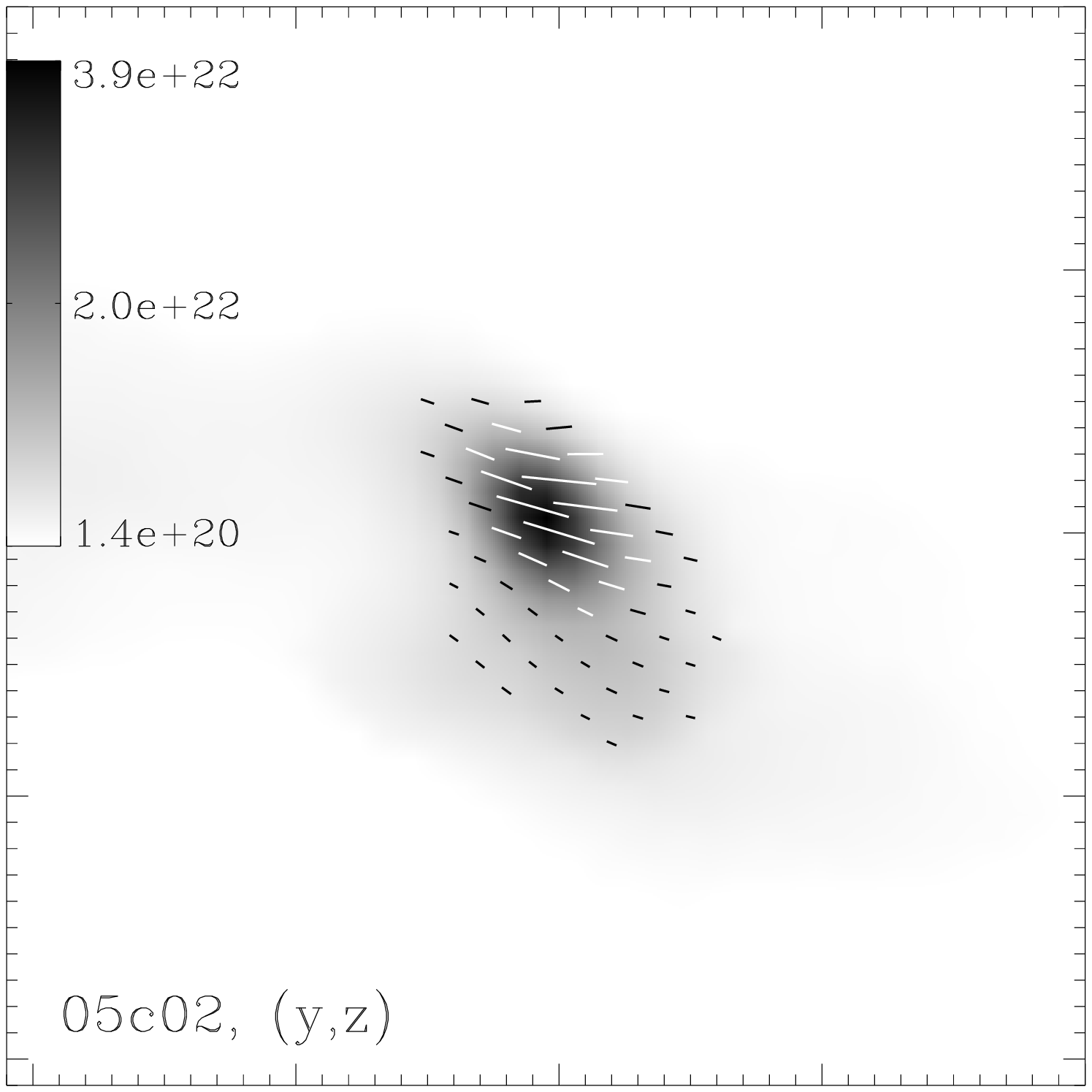}
  \hfill
  \includegraphics[width=0.32\textwidth]{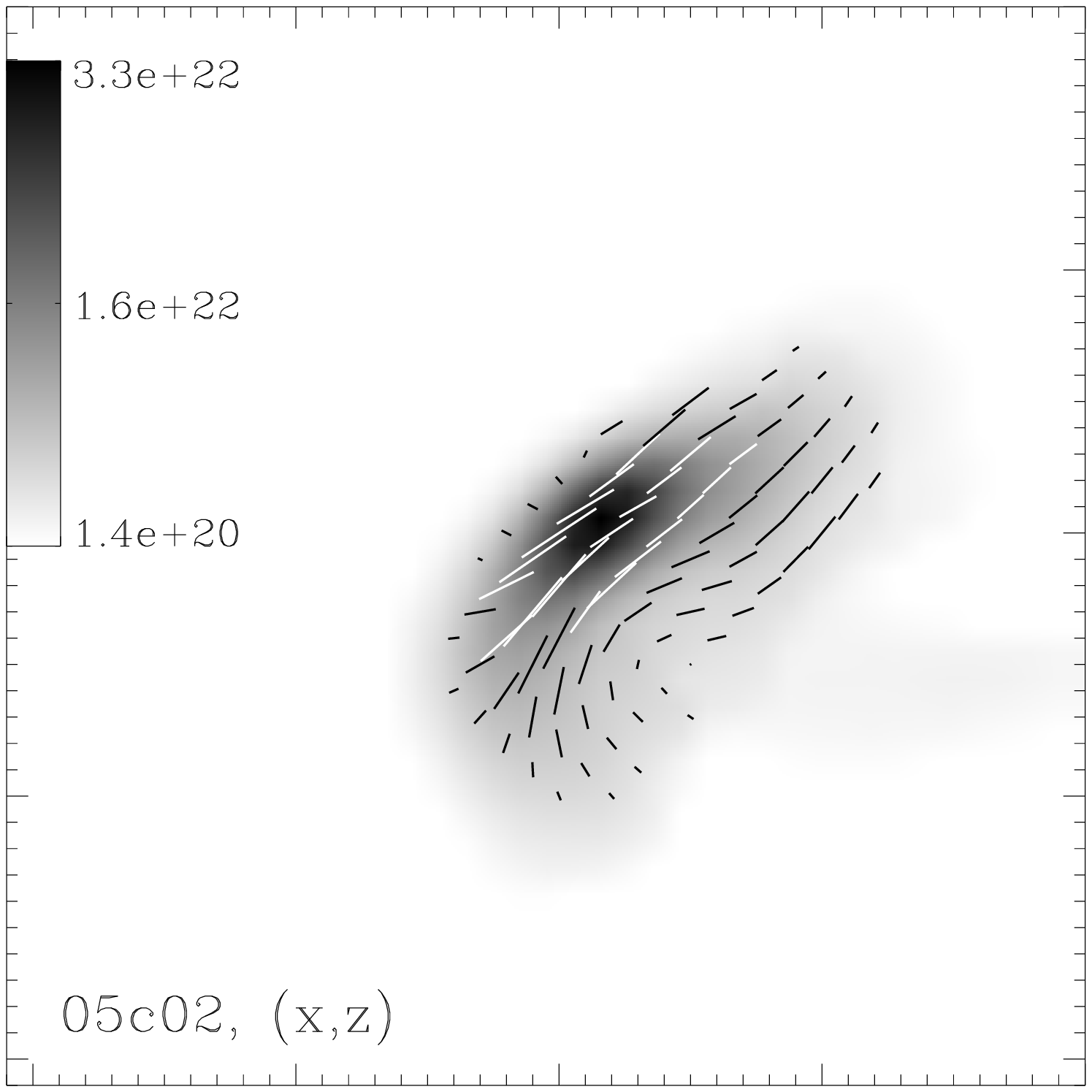}
  \hfill
  \includegraphics[width=0.32\textwidth]{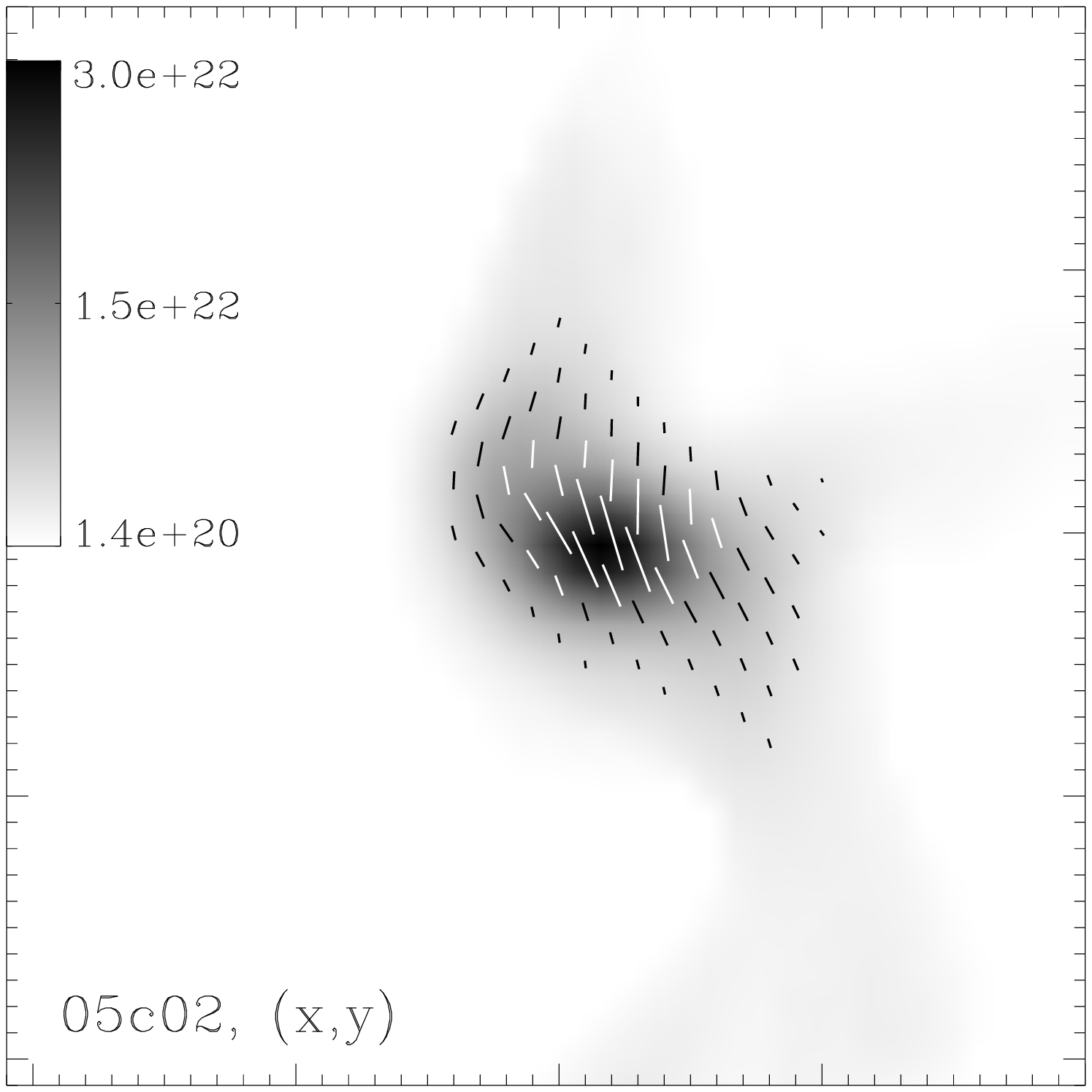}
  \caption{\label{f:coremaps}
           (a) Polarization map looking along the mean flux direction.
           (b,c) Polarization map looking perpendicular to the mean flux
           direction denoted in Fig. \ref{f:core3d}. One panel measures
           approximately $0.25$pc across. The color bars give the column 
           density in particles cm$^{-2}$. Every second polarization 
           vector is shown (and used for the field estimate), resulting in
           a spatial ``resolution'' of $0.012$pc.}
\end{figure*}

\begin{figure*}
  \includegraphics[width=0.32\textwidth]{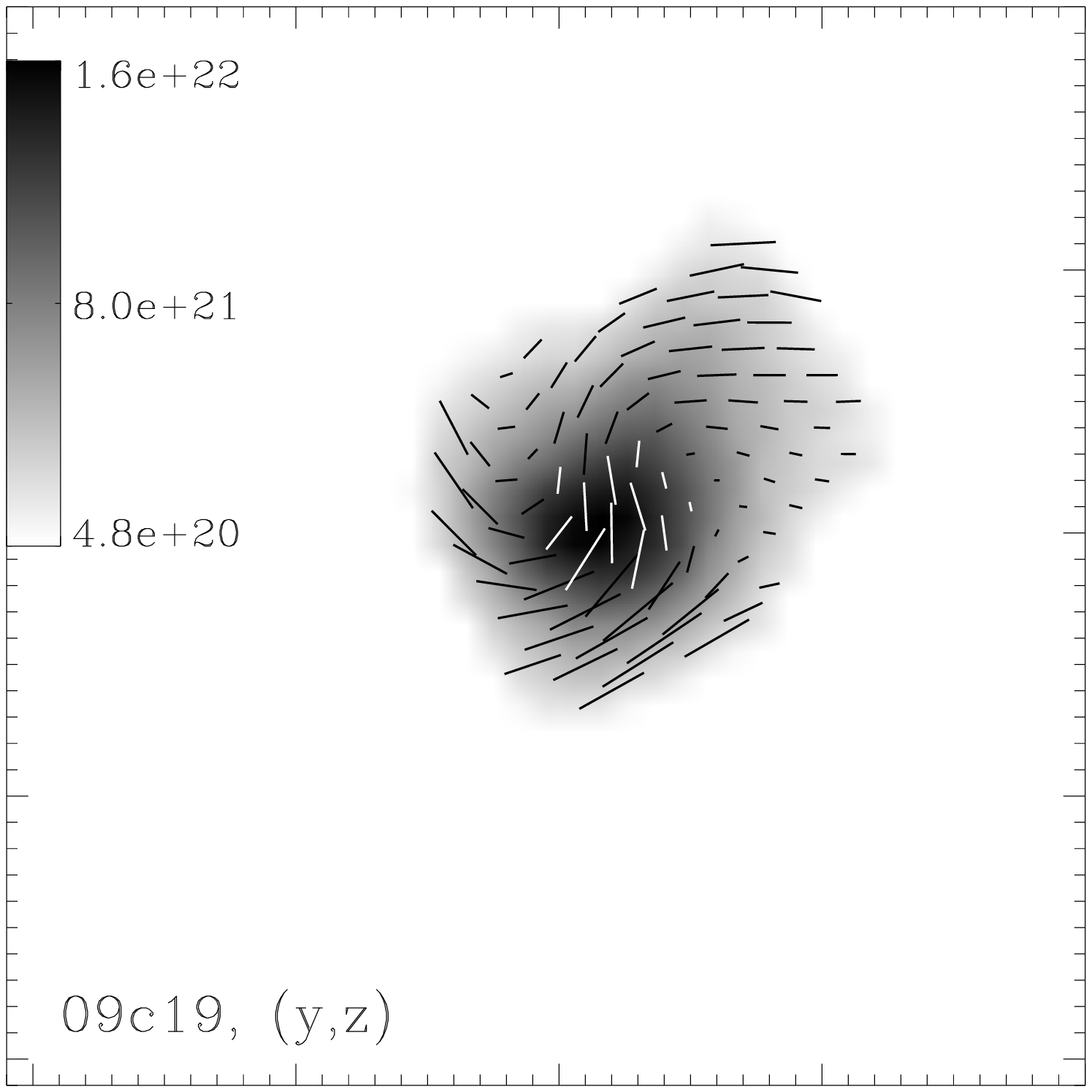}
  \hfill
  \includegraphics[width=0.32\textwidth]{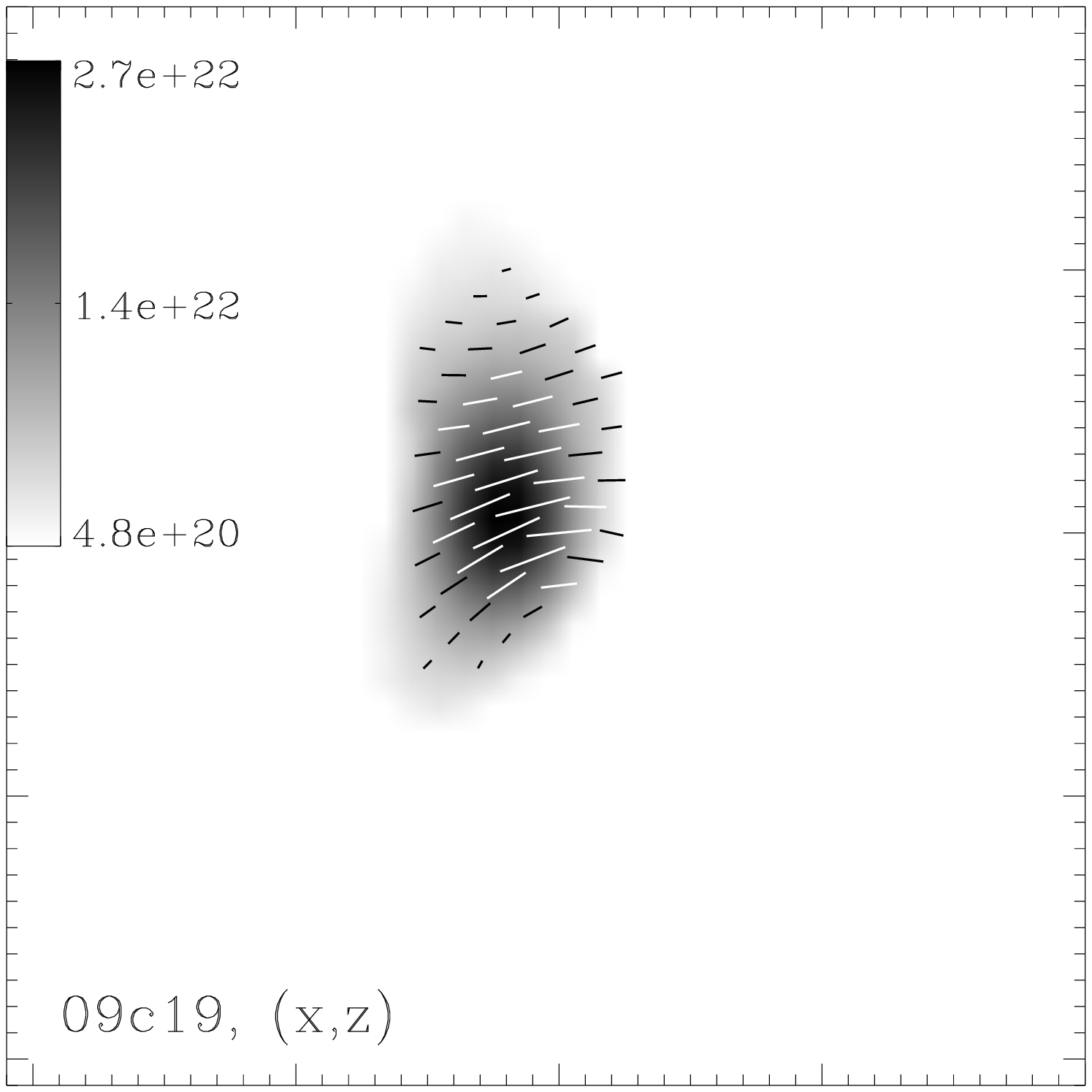}
  \hfill
  \includegraphics[width=0.32\textwidth]{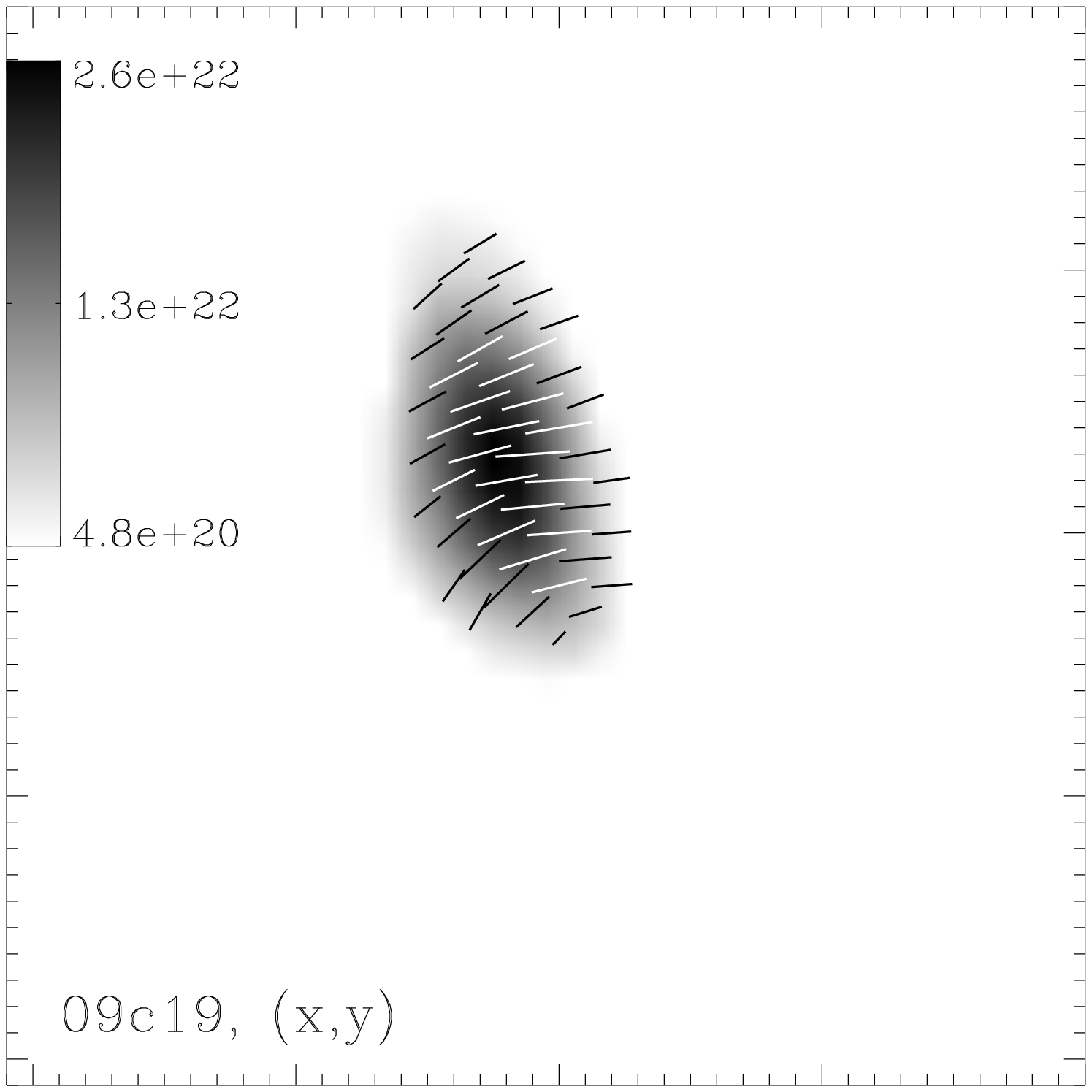}
  \caption{\label{f:corefig22}
           (a) Polarization map looking along the mean flux direction for
           the core shown in Fig. 22 of \citet{LNM2004}. Note the spiral
           polarization pattern induced by the core's rotation.
           (b,c) Polarization map looking perpendicular to the mean flux
           direction. One panel measures
           approximately $0.25$pc across. The color bars give the column 
           density in particles cm$^{-2}$.  Every second polarization vector
           is shown (and used for the field estimate), resulting in
           a spatial ``resolution'' of $0.012$pc.}
\end{figure*}

In order to establish the reliability of the CF-method for cores,
we generated polarization maps after rotating the cores
such that the line of sight is either parallel to the mean flux
direction (the ``worst case'' for the CF-method, since then the 
necessary mean field component would be expected to vanish) or
perpendicular to the mean flux direction (``best'' case).
An example of the three-dimensional core structure and the resulting
polarization maps is given in Figures~\ref{f:core3d} and \ref{f:coremaps}.
The depicted core is a typical example in the sense that it combines
very ordered with turbulent field structure. 
Figure~\ref{f:corefig22} displays the three polarization maps -- again
seen along the mean flux direction and perpendicular to it -- for
the core shown in Figure 22 of \citet{LNM2004}. The oblate core is permeated
by field lines oriented in parallel to its rotation axis. The field lines are
slowly wound up, resulting in a spiral pattern when viewing the core along its
rotation axis (left panel of Fig.~\ref{f:corefig22}). For the lines of sight
perpendicular to the rotation axis, the rotation pattern does not emerge (center
and right panel).

\subsection{Performance of the CF-method in cores\label{ss:cfperformace}}

\begin{figure}
  \plotone{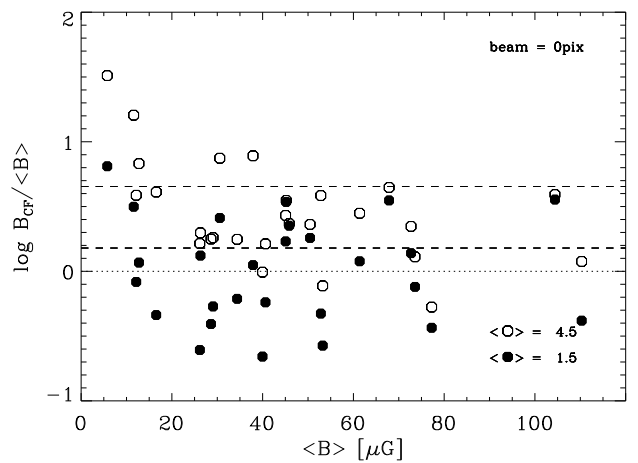}
  \caption{\label{f:cfperformance}Reliability $a = B_{CF}/\langle B\rangle$
  versus the mean field in the core $\langle B\rangle$ for all resolved cores
  and lines of sight. Open symbols: angular dispersion selected for
  $\Delta\theta < 25^{\circ}$. Filled symbols: angular dispersion selected for
  $\Delta\theta < 85^{\circ}$. The dashed lines denote the averages, whose values are
  given in the plot. No smoothing applied. 
  Clearly, selecting for $25^{\circ}$ leads to a systematic overestimate, while
  using the full information leads to an overestimate by approximately $2$
  (lower dashed line).}
\end{figure}

Figure~\ref{f:cfperformance} summarizes the reliability estimates for the
CF-method applied to the resolved model cores (i.e. the ones without X's in 
Table~\ref{t:cores}). For this plot, all three lines 
of sight were used. First, we note that using the small angle approximation
(open symbols) leads to a systematic overestimate of the field strength by 
a factor of more than $4$. This is not surprising since we discard physically 
relevant information about the field structure (and thus strength) when 
selecting for small angles. However, we cannot restrict the line of sight
velocity dispersion in the same way. On the other hand, using 
equation~(\ref{e:cfextend}) (filled symbols) leads to a moderate overestimate
of the field by a factor of approximately two\footnote{The selection for
$\Delta\theta < 85^{\circ}$ is only applied to prevent the tangent from blowing up.}. 
This is not inconsistent with
the findings of \citet{OSG2001}, since they selected for their 
whole domain and thus their results were not affected by small-number
statistics. For large-number statistics, weak fields will lead to a wide angle 
dispersion, meaning that even when selecting, most of the angles are found at 
the largest possible values, thus indicating a weak field.
\begin{figure}
  \plotone{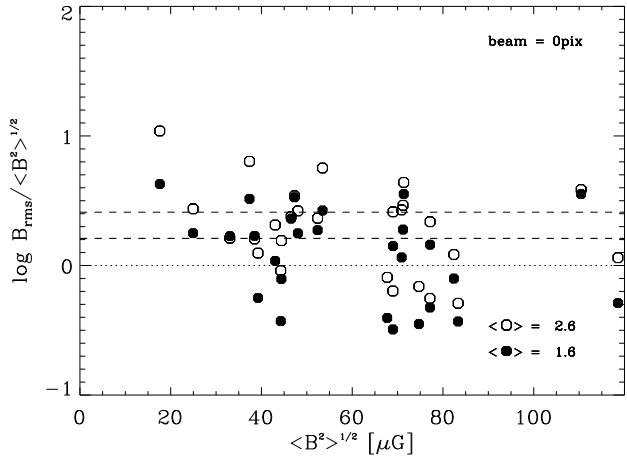}
  \caption{\label{f:zwperformance}Reliability $a = B_{rms}/\langle B^2\rangle^{1/2}$
  versus the rms mean field in the core $\langle B^2\rangle^{1/2}$ for all resolved cores
  and lines of sight. Open symbols: angular dispersion selected for
  $\Delta\theta < 25^{\circ}$. Filled symbols: angular dispersion selected for
  $\Delta\theta < 85^{\circ}$. The dashed lines denote the averages, whose values are
  given in the plot. No smoothing applied.}
\end{figure}

The small number of strong-field cores is conspicuous. This is simply a consequence
of low initial field strength, which renders the turbulent box supercritical
by a factor of $8$. On the one hand, this is inconsistent with the observations, on 
the other hand it allows us to test the CF-method under severer conditions 
than it would probably meet in observational data.

The modified method to estimate the rms field in the core gives -- within the scatter
-- the same estimated field (Fig.~\ref{f:zwperformance}). 
For smooth fields the method reverts to the original version, while for unordered
fields, the method emphasizes the angular variations (see HZMLN for a 
discussion). The scatter of the estimates is slightly reduced. Note that 
on the whole the rms field is somewhat larger than the mean field, indicating a 
non-uniform field component in the cores.

Figures~\ref{f:cfperformance}, \ref{f:zwperformance} and all subsequent ones
show cores extracted at different times during the simulation. Although we thus do not
expect to be affected by selection effects that could result from chosing a 
special time instant during the
simulation, there is of course a selection effect in the sense that we extract cores
at comparable evolutionary stages, i.e. cores that are already self-gravitating, but
still numerically resolved. As long as the cores are resolved, a correlation
between reliability and global evolution time (last column in Table~\ref{t:cores})
cannot be seen.

Applying equation~(\ref{e:cfextend}) and preventing the angular dispersion
from blowing up by selecting for angle differences $\Delta\theta < 85^{\circ}$
improves the reliability of the estimates in these models.

\subsection{Effects of limited observational and numerical resolution\label{ss:limitres}}
\begin{figure}[h]
  \plotone{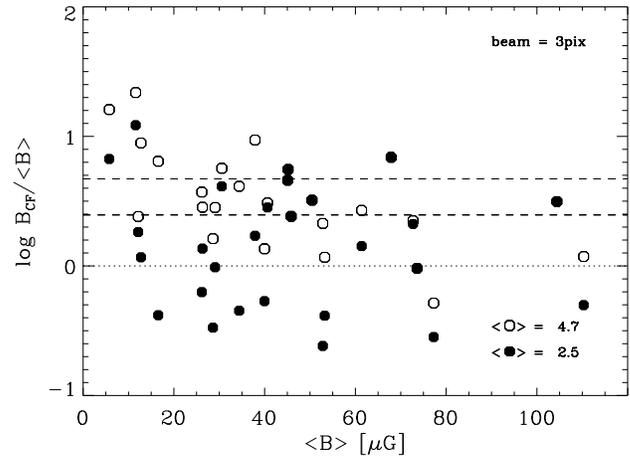}
  \caption{\label{f:cfperformancesmooth}Reliability $a = B_{CF}/\langle B\rangle$
  versus the mean field in the core $\langle B\rangle$ for all resolved cores
  and lines of sight at reduced resolution (map smoothed with a Gaussian
  of $\sigma=3$pix). Open symbols: angular dispersion selected for
  $\Delta\theta < 25^{\circ}$. Filled symbols: angular dispersion selected for
  $\Delta\theta < 85^{\circ}$. The dashed lines denote the averages, whose values are given
  in the plot.}
\end{figure}

As HZMLN showed for extended regions, limited telescope resolution 
(finite beam width) leads to a systematic overestimate, because the field variations
are averaged out, making the field look smoother, i.e. in terms of the
method, stronger. We can observe a similar effect in Figure~\ref{f:cfperformancesmooth},
compared to Figure~\ref{f:cfperformance}. 
Degrading the resolution by roughly a factor
of $3$ leads to an increased overestimate. The estimates selected for 
$\Delta \theta < 25^{\circ}$ are less affected by the degraded resolution, and both 
selected sets of estimates agree for strong fields, as is to be expected.

It might be interesting to note that the ambipolar diffusion scale 
\begin{equation}
  L_{AD} = \frac{\lAD}{v_{rms}} = 1.97\times 10^{31}
           \frac{B^2}{x_i\,n_n^2\,v_{rms}}
  \label{e:decouplingscale}
\end{equation}
is on the order of $0.01$pc for typical core parameters. 
In equation~(\ref{e:decouplingscale}) the field strength $B$, the neutral particle 
density $n_n$ and the velocity dispersion $v_{rms}$ are given in cgs-units. 
The ionization 
fraction is denoted by $x_i$.

\begin{figure}[h]
  \plotone{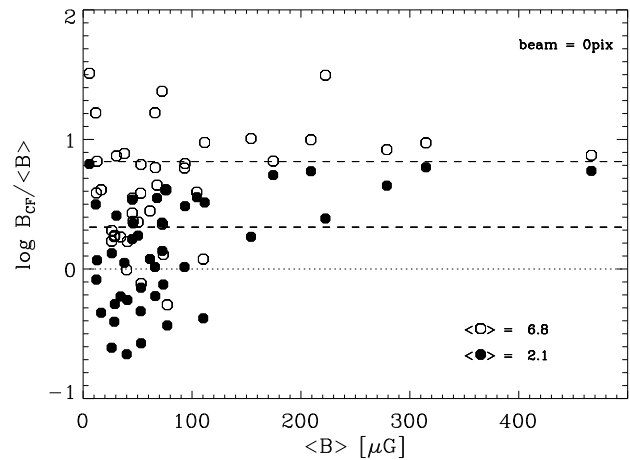}
  \caption{\label{f:cfperformanceunres}Reliability $a = B_{CF}/\langle B\rangle$
  versus the mean field in the core $\langle B\rangle$ for all sample 
  cores (including unresolved ones, see Tab.~\ref{t:cores})
  and lines of sight. Open symbols: angular dispersion selected for
  $\Delta\theta < 25^{\circ}$. Filled symbols: angular dispersion selected for
  $\Delta\theta < 85^{\circ}$. The dashed lines denote the averages, whose values are given
  in the plot.}
\end{figure}

While Figure~\ref{f:cfperformancesmooth} demonstrated the effect of degraded telescope
resolution, we exemplify the influence of under-resolved cores on the overall result
in Figure~\ref{f:cfperformanceunres}.
The effect is obvious: The peak densities in those cores start to grow unhampered
once they overstep the resolution criterion. The peak is concentrated in the central
cell, so that the field is not able to follow the density evolution \citep{HMK2001}. 
Because of the over-large densities, the CF-method systematically overestimates
the true field strength, a purely numerical effect in this case. Note that the
overestimates correlate with the strongest fields and unresolved cores.

The insufficient numerical resolution leads to a diffusion of the field out
of the dense region, qualitatively similar to ambipolar diffusion.
While there is no basis on which to draw any conclusions from these
under-resolved cores (especially since the numerical diffusion cannot be 
controlled, and can behave quite differently from its physical counterpart),
a similar effect would be expected in the observations {\em if}
the density tracer is a neutral species.

\subsection{CF Method vs. Zeeman Effect\label{ss:cfvszeeman}}

\begin{figure}
  \plotone{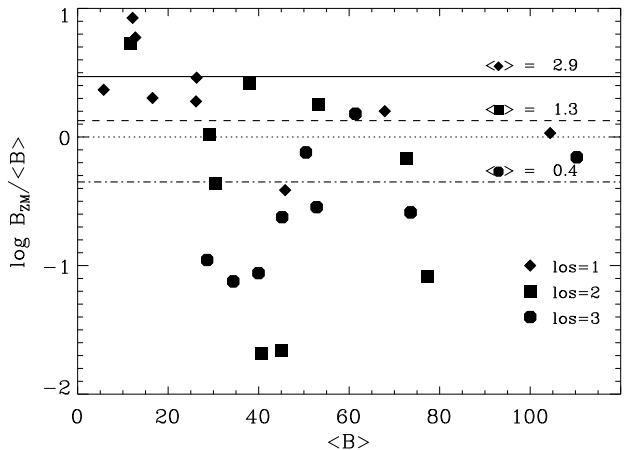}
  \caption{\label{f:zeeman}Reliability for the Zeeman estimates 
  $B_{ZM}/\langle B\rangle$ against the mean field in the core. 
  Averages broken up according to line of sights are given in the plot,
  and los=1 means integration along the mean flux direction (as in Fig.~\ref{f:coremaps}).}
\end{figure}

If the magnetic field is not uniform, Zeeman measurements -- realized in our models 
via equation~(\ref{e:zeeman}) -- are expected to systematically underestimate the
field strength because of cancellation by field reversals along the line of sight.
At first glance, this expectation is not met (Fig.~\ref{f:zeeman}). 
The Zeeman-effect seems to overestimate the field roughly by the same 
amount as the CF-method does. However, distinguishing between the lines of sight
clarifies the picture. Zeeman estimates taken along the mean flux direction -- denoted
by ``los=1'' in Figure~\ref{f:zeeman} -- yield larger values than those taken 
perpendicularly to the mean flux direction (``los=2'' and ``los=3''). 
Since the line of sight integration is density-weighted, fields at locations with higher
densities tend to be emphasized. Because of flux-freezing, this generally means that 
there is a tendency to higher field strength estimates in cores with large density contrasts,
leading to a systematic overestimate for such cores. Thus the estimates along the mean
flux direction are dominated by the strongly weighted central field values, while the
estimates perpendicular to the mean flux direction ``see'' the perturbed component
around it, leading to cancellation. This of course holds only as long as the emissivity
is (approximately) proportional to the density, as assumed in equation~(\ref{e:zeeman}).

\begin{figure}
  \plotone{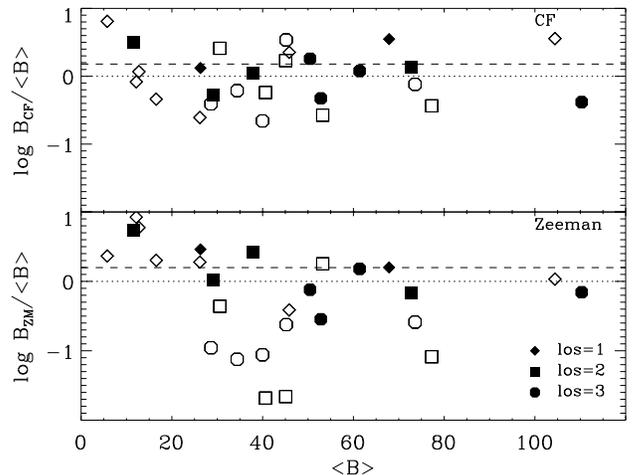}
  \caption{\label{f:cfvszeeman}Reliability for the CF-estimates as in 
  Figure~\ref{f:cfperformance} (upper panel) and for the ``Zeeman'' estimates 
  $B_{ZM}/\langle B\rangle$ (lower panel) against the mean field in the core. 
  The dashed line gives the mean reliability. los=1 means integration along the
  mean flux direction (as in Fig.~\ref{f:coremaps}). Filled symbols denote estimates
  for which $B_{CF}$ and $B_{ZM}$ agree within a factor of $3$, open symbols stand
  for the remaining estimates.}
\end{figure}

Comparing the CF and Zeeman estimates in Figure~\ref{f:cfvszeeman},
the picture is somewhat less clear for the CF-method. There is a slight tendency to 
a larger scatter for estimates along the mean flux direction (upper panel of 
Fig.~\ref{f:cfvszeeman}, diamonds), as one would expect, since the method is expected
to work most reliably for a strong mean field component in the plane of sky. 
Unfortunately, this is a reliability criterion gathered in ``hindsight'', i.e. it is
not possible to apply it to observational data. 

Instead, we might be tempted to argue that for cases where the CF-method and the Zeeman
measurements yield approximately the same results, the field value should be well 
determined. Filled symbols in Figure~\ref{f:cfvszeeman} denote estimates for which 
$B_{CF}$ and $B_{ZM}$ agree within a factor of three. The scatter is reduced for both
the CF-method and the Zeeman-measurement, although we should note that the estimates could
still be off by a factor of approximately five.
It might be interesting to note that the overall scatter of the Zeeman measurements around the 
``true'' field value is actually larger than that of the CF-estimates (Fig.~\ref{f:cfvszeeman}).

\subsection{Effects on derived criticality of cores\label{ss:criticality}}

\begin{figure}
  \plotone{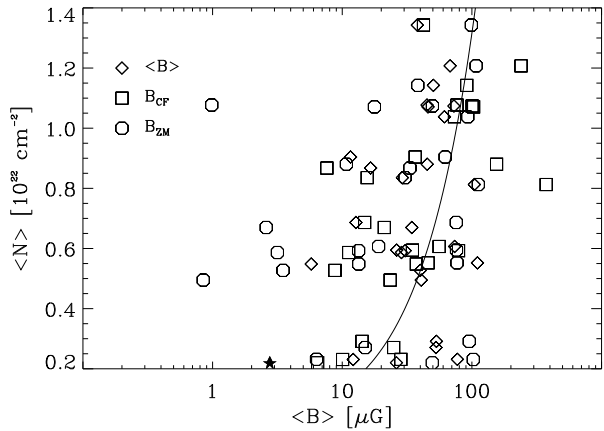}
  \caption{\label{f:criticality}Column density $N$ against magnetic field
           strength $B$ to estimate criticality of cores. The line gives
           $\lambda\equiv 1$ (see eq.~[1] of \citet{CNW2004}). The symbols denote
           how $N(B)$ is determined. Diamonds: directly from simulations, 
           squares: CF-Method, circles: Zeeman effect. Only resolved cores are 
           shown (see Tab.~\ref{t:cores}), selected for $\Delta \theta < 85^{\circ}$ 
           in the case of the CF-estimates. Different symbols on constant lines of $N$ belong
           to one line of sight of one core. The star denotes $N(B)$ for the full 
           simulation domain (i.e. $27$ pc$^3$).}
\end{figure}

Figure~\ref{f:criticality} demonstrates the scatter of the magnetic field estimates in
terms of the criticality parameter $\lambda$ as defined by \citet{CNW2004}, eq.(1), namely
\begin{equation}
  \lambda = 7.6\times 10^{-21}\frac{N(\mbox{H}_2)}{B},
\end{equation}
where the magnetic field $B$ is given in $\mu$G and the H$_2$ column density in 
cm$^{-2}$. 

The tendency of the CF-method (squares in Fig.~\ref{f:criticality})
to generally overestimate the field strength renders the core ensemble slightly more 
subcritical than expected from the numerical values (diamonds in Fig.~\ref{f:criticality}).
There are a few subcritical numerical data points (diamonds). 
These are strongly elongated filaments with fields aligned in parallel, indicating 
field line stretching. Although the numerical values already show a considerable scatter because
of dynamical effects, the uncertainties in the CF-estimates as well as in the Zeeman-estimates
emphasize that field estimates of single cores allow only a limited conclusion about the 
physical state of the core.
 
%
%
\section{Discussion}\label{s:conclusions}

Zeeman measurements and the CF-method are currently the most viable ways
to estimate the magnetic field strength in the dense molecular medium. 
Since the two methods are measuring different components of the field,
they complement each other to some extent. The CF-method rests on the assumption
of equipartition between turbulent kinetic and turbulent magnetic energy in
the medium, which is not necessarily guaranteed in regions dominated 
by e.g gravity or rotation, such as protostellar cores.

We tested with the help of model cores formed in a numerical simulation of
self-gravitating magnetized turbulence how accurately the CF-method and 
Zeeman measurements can estimate the magnetic field strength in such cores. We find
\begin{enumerate}
  \item that the CF-method is on the average 
  surprisingly robust even in regions not dominated by turbulence. 
  The previous overestimation factor of $2$ is still valid approximately
  for the model cores on the average, {\em if} the full angle dispersion information 
  is used. Single measurements can be off by a factor of up to $7$ in both
  directions.
  \item that applying the small-angle approximation in the CF-method 
  leads to a systematic overestimate of $\langle B\rangle$.
  \item that the ``Zeeman'' estimates of our cores can lead to an 
  {\em overestimate} for integrations along the mean flux direction, 
  resulting from the emphasis on strong fields from the density-weighted line of sight
  integration. For integrations perpendicular to the mean flux direction, the Zeeman
  measurements tend to underestimate the field strength. Thus, Zeeman measurements
  cannot generally be regarded as a lower limit on the field strength.
\end{enumerate}

A major shortcoming of our model is its low field strength, which renders its regime
less magnetized than observed in molecular cores. While this is definitely a limitation,
we see it also as an advantage since the CF-method is expected to work
more reliably for stronger field strengths (HZMLN). Obviously, further limitations of 
this study stem from the limited resolution within the cores.

Taking the emissivity proportional to the density (eq.~\ref{e:zeeman}) in the Zeeman
estimates does not account for the fact that the tracer species do not necessarily
contribute proportionally to gas density. In that sense, the overestimate inferred
for the Zeeman measurements from our models could be regarded as an upper limit. 
We did not address the problem of how to convert tracer column densities into H$_2$ 
densities, which is very likely another major source of uncertainty in the CF-method.
Similarly, we left out a discussion of the velocity tracers. 

We conclude that single magnetic field estimates in molecular cores should
be regarded with caution because of the substantial scatter. The CF-method works
reliably on average and should be combined with other types of estimates. 

\acknowledgements
This study relies on the $512^3$ simulation \citep{LNM2004} calculated on
the SGI Origin 2000 of the National Center of Supercomputing Applications
at the University of Illinois at Urbana-Champaign. 
We thank G. Birk, R.~M. Crutcher, J.~S. Greaves, A. Lazarian, B.~C. Matthews and 
E.~G. Zweibel for enlightening discussions. The helpful comments by 
M.-M. Mac Low and E.~G. Zweibel on the draft are very much appreciated. 
F.H. is grateful for the support by a Feodor-Lynen grant from the
Alexander von Humboldt Foundation. 
The work was supported by NSF grants AST-0098701 and AST-0328821.
The data were analyzed at the NCSA and 
on the local PC cluster at UW-Madison, built by S. Jansen.

%
%

\end{document}